\documentclass[twocolumn]{article}

\usepackage{preprint}

\usepackage[T1]{fontenc}

\usepackage{amsmath}
\usepackage{amssymb}
\usepackage{siunitx}

\usepackage{color}
\usepackage{graphicx}
\usepackage{tikz}

\usepackage{multirow}
\usepackage{multicol}
\usepackage{float}
\usepackage{subcaption}

\usepackage{longtable}

\usepackage[inline]{enumitem}

\usepackage[numbers]{natbib}

\usepackage{xwatermark}
\newwatermark[firstpage,color=gray!60,angle=90,scale=0.32, xpos=3.9in,ypos=0]{\href{https://doi.org/}{\color{gray}{Publication DOI}}}
\newwatermark[firstpage,color=gray!90,angle=0,scale=0.28, xpos=0in,ypos=-5in]{Preprint submitted to Journal of Crystal Growth}

\usepackage[colorlinks = true,
            linkcolor = blue,
            urlcolor  = blue,
            citecolor = blue,
			anchorcolor = black]{hyperref}
			
\usepackage{abstract}

\newcommand{\Real}{\mathrm{Re}}

\title{Inverse response behaviour in the bright ring radius measurement of the Czochralski process II: Mitigation by control}

\usepackage{authblk}

\author[1]{Halima Zahra Bukhari}
\author[1]{Morten Hovd}
\author[2\thanks{Corresponding author: \texttt{jan.winkler@tu-dresden.de}}]{Jan Winkler}

\affil[1]{Department of Engineering Cybernetics, NTNU, Trondheim, Norway}
\affil[2]{Institute of Control Theory, Faculty of Electrical \& Computer Engineering, TU Dresden, Germany}

\begin{document}

\twocolumn[ 
\begin{@twocolumnfalse} 

\maketitle

\begin{abstract}
This is the second part of a two-article series investigating the presence of an
inverse response in the measurement of the ingot radius in the Czochralski
process for monocrystalline silicon production, when the ingot radius is deduced
from a camera image of the bright ring at the meniscus connecting the solid
crystal to the silicon melt. Such inverse responses are known to pose a
fundamental limitation in achievable control performance. However, for the
bright ring radius measurement, the inverse response is an artefact of the
measurement technique and does not appear in the physical variable one wants to
control (the actual crystal radius). The present article addresses control for
the mitigation of the inverse response behaviour, using a combination of
parallel compensation and feedback control. The proposed design is validated
against simulations where the production process is subjected to temperature
disturbances.
\end{abstract}

\vspace{0.35cm}

  \end{@twocolumnfalse} 
] 
\saythanks

\section{Introduction}
The Czochralski (Cz) process is the industrially dominant process for the production
of monocrystalline silicon ingots. The ingots produced are cut into thin wafers,
which are the basis for the production of photovoltaic cells and computer chips. Any
variation in the ingot radius will increase the material waste in the subsequent
production steps and might also initiate defects in the crystal structure during
growth. Good control of the ingot radius is, therefore, important. Unfortunately,
the quantity measured for feedback control, the \emph{bright ring radius}, is
affected by an \emph{inverse response} behaviour \citep{gevelber1994dynamics}.
That is, when the crystal pulling speed is increased, the radius measurement at
first increases and thereafter decreases. However, in general, the steady-state effect of the increased pulling speed is a decreased radius measurement, and vice versa. The term
\emph{inverse response} refers to the phenomenon when the initial response of
the controlled variable is in the opposite direction of the steady-state
response. Such inverse responses are one appearance of what in control parlance
is termed \emph{non-minimum phase behaviour}, and causes fundamental limitations
in achievable control performance \citep{skogestad2007multivariable}. This paper
presents a control design approach, where a conventional PID feedback controller
is combined with a so-called \emph{parallel compensator} to circumvent the
limitations on feedback control caused by the non-minimum phase dynamics of the
system.

\subsection{Paper Organization}
A description of the Cz process is given in the first part of this article
series \cite{BUKHARI2020A}, where the model relevant for the control of crystal
growth is developed. The model combines rigorous descriptions of the meniscus
shape and the camera-based radius measurement with simplified temperature
dynamics. The analysis of the developed model verifies the presence of the
inverse response from pulling speed to radius measurement. Based on this, the
present paper is structured as follows: Section\,\ref{sec:GrowthModel} studies the
dynamics of the crystal radius control problem, elucidating the requirements and
limitations in relevance to the control design. For this purpose, the heat
transfer from the heaters to the melt is of minor significance, and a reduced
model with a constant melt temperature is used. This model is linearized in Section\,\ref{sec:Linearization} so that in Section\,
\ref{compensator_controller} the linear controller and parallel compensator
can be designed. In Section\,\ref{sec:temp_model}, the controller design is verified in
simulations involving the overall system model from \cite{BUKHARI2020A}, and it
is shown that the controller achieves stable control also of the nonlinear
dynamics, as well as good suppression of disturbances entering through the temperature dynamics.

\section{Cz growth model and inverse response behaviour}\label{sec:GrowthModel}
In the first part \cite{BUKHARI2020A} of this article series, two models for the
crystal growth and radius dynamics are developed. These two models differ in
how the heat transfer from the melt to the melt/crystal interface are modelled:
\begin{enumerate}[label=\Roman*., widest=II,labelindent=1ex,leftmargin=*]
\item In this model, the heat transfer across the meniscus is modelled based on conductive
flows, whereas the transfer of heat from the melt bulk to the meniscus is
governed dominantly by convective flows.
\item Here, the overall heat transfer from the bulk of the melt to the
melt/ crystal interface is dominated primarily by the convective heat flow.
\end{enumerate}

Of course, the two models are clear simplifications of reality, with the actual heat transfer across the meniscus likely to be somewhere between pure conduction and pure convection.

As a consequence of these assumptions for \textbf{Model I}, the overall heat
transfer to the interface depends on the meniscus height (cf.\,Eq.\,(8) in part
I of this article series), while in \textbf{Model II}, the heat transfer to the
interface is independent of the meniscus height (cf.\,Eq.\,(12) of part I). The
same holds for the growth rate since it directly depends on this heat transfer
(cf.\,Eq.\,(1e) in part I).

\begin{figure}[ht!]
    \centering
    \includegraphics[width=0.8\linewidth]{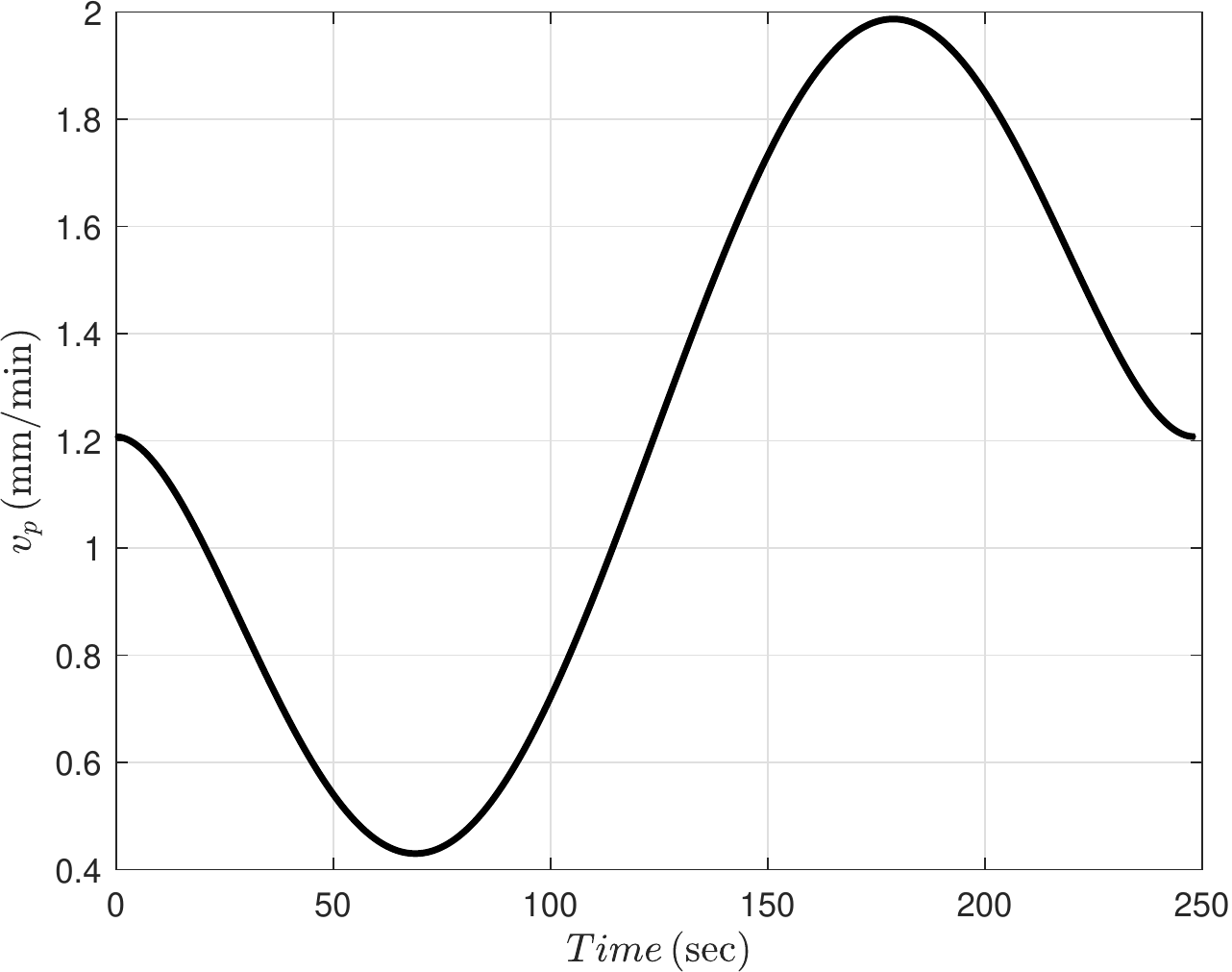}
    \caption{Applied pull speed profile}
    \label{fig:vp_input}
\end{figure}

\begin{figure}[ht!]
    \centering
    \includegraphics[width=1.0\linewidth]{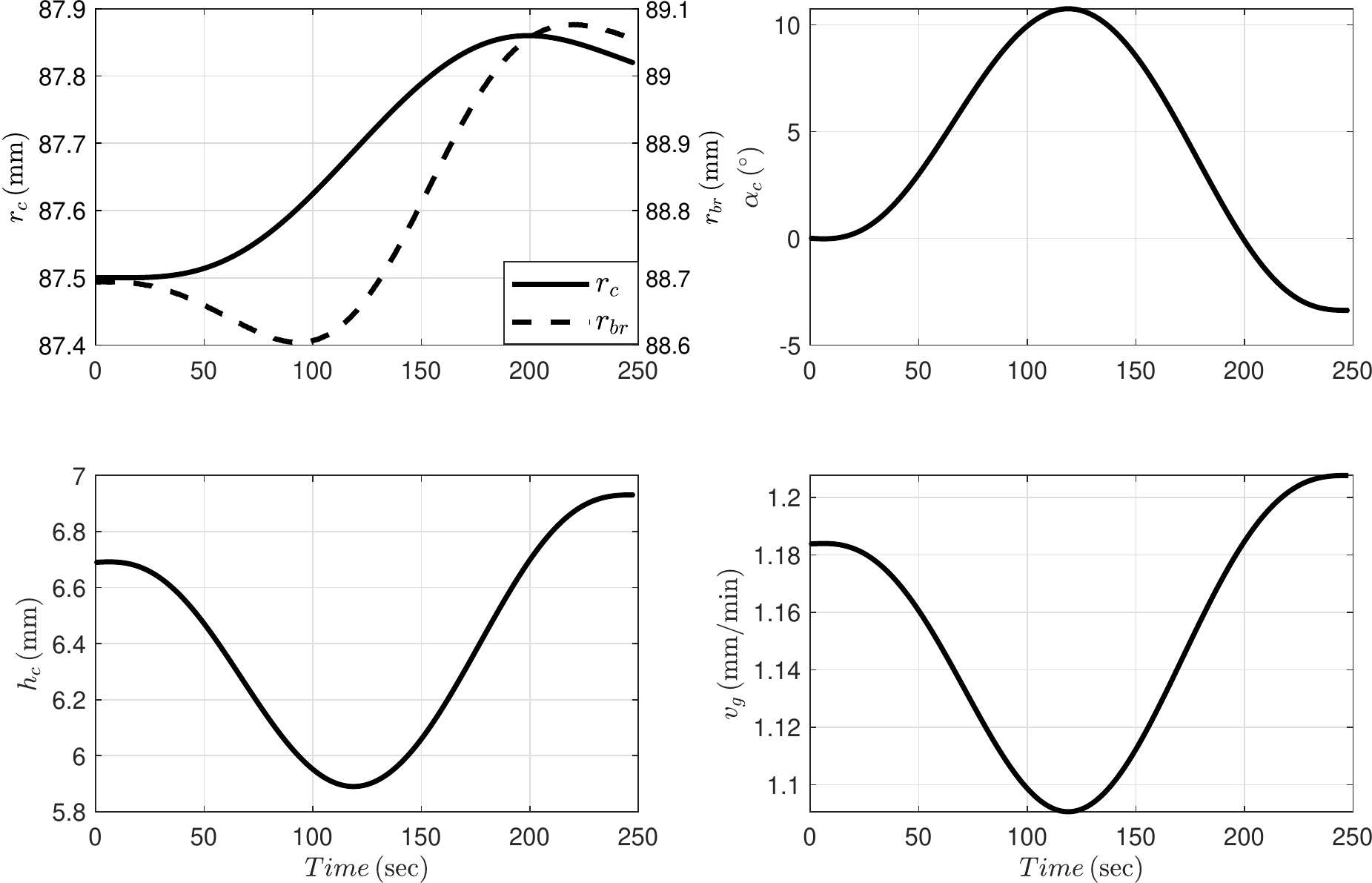}
    \caption{Nonlinear plant response for \textbf{Model I} subjected to the applied input shown in Fig.\,\ref{fig:vp_input}, crystal radius ($r_{c}$), bright ring radius ($r_{br}$) and crystal growth angle ($\alpha_c$) in top-left and right panes, respectively, the meniscus height ($h_{c}$) and the growth rate ($v_{g}$) in the bottom-left and right panes, respectively.}
    \label{fig:non_lin_conduction}
\end{figure}

\begin{figure}[ht!]
	\centering
	\includegraphics[width=1.0\linewidth]{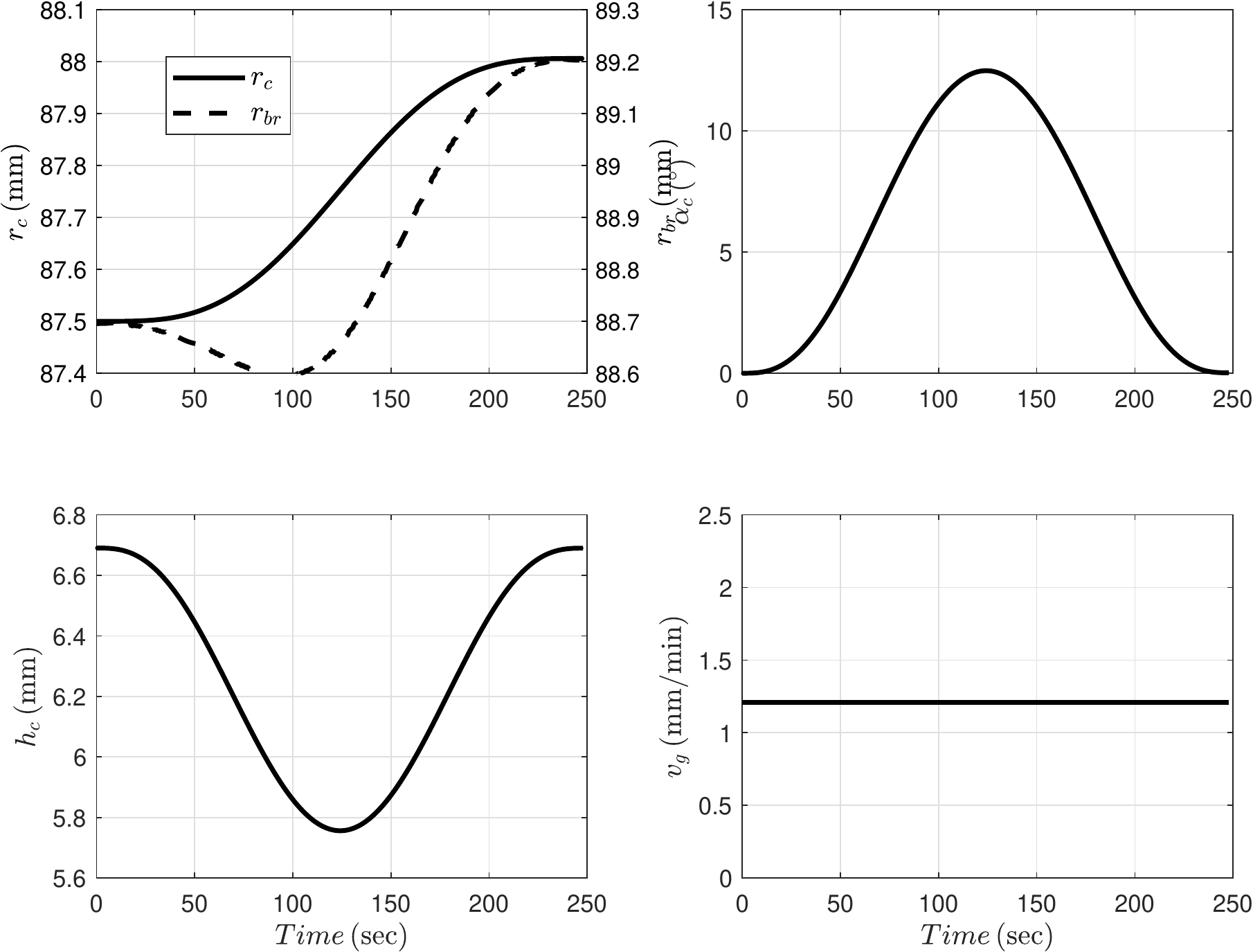}
	\caption{Nonlinear plant response for \textbf{Model II} subjected to the applied input shown in Fig.\,\ref{fig:vp_input}, crystal radius ($r_{c}$) and bright ring radius ($r_{br}$), crystal growth angle ($\alpha_c$) in top-left and right panes respectively, the meniscus height ($h_c$) and the growth rate ($v_g$) in the bottom-left and right panes, respectively.}
	\label{fig:non_lin_convection}
\end{figure}

To illustrate the difference between the two models, both models are simulated with the smooth profile for the crystal pulling speed shown in Fig.\,\ref{fig:vp_input} while the heater power is kept constant. The resulting
responses of the two models are depicted in Figs.\,\ref{fig:non_lin_conduction}
and \ref{fig:non_lin_convection} for models I and II, respectively.


It is apparent that a decrease in pull speed causes the crystal radius to
increase. However, the measurement of the bright ring radius, (which is used as an estimate
 of the crystal radius) initially moves in a direction opposite to that of
the actual crystal radius, thereby confirming the presence of inverse response. Moreover, the heat flux into the interface based
on convective heat transfer (Model II) yields a constant growth rate $v_g$, whereas it
varies with height for the case with pure conductive heat transfer across the
meniscus (Model I).

\section{Linearized model} \label{sec:Linearization}
For the approach presented in this paper, it is necessary to linearize the nonlinear
model around a steady-state crystal radius. While the equations describing the
meniscus dynamics and temperature effects are given in closed form, the
dependency of the measured bright ring radius on the crystal radius and the
pulling speed is not. Hence, a full-model numerical perturbation linearization
is carried out around a steady-state crystal radius of \SI{87.5}{\mm} and a
pulling speed of \SI{1.2}{\mm\per\minute}. The obtained linear ordinary
differential equations are then transformed into a corresponding transfer
function $G(s)$ in the complex-valued Laplace domain with the complex-valued
Laplace variable $s$ (cf.\,Section \ref{sec:control:models}).
This transfer function describes the linear response of the bright ring radius
$R_{br}$ to changes in the crystal pulling speed $V_{p}$, where the uppercase written variables indicate being in the Laplace domain.

\begin{figure}[ht]
	\centering
	\includegraphics[width=0.75\linewidth]{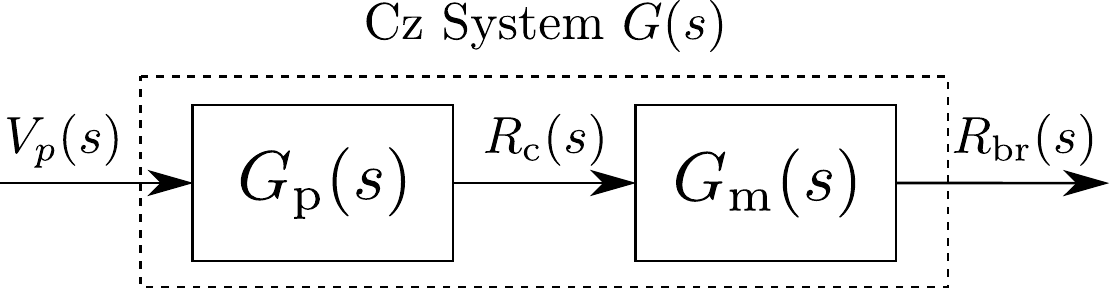}
	\caption{Series connection of the transfer functions describing the pulling speed - crystal radius dynamics $G_p(s)$ and the measurement dynamics $G_m(s)$.}
	\label{fig:series-connection}
\end{figure}

It will prove useful in this paper to formulate this transfer function as a series connection of the process dynamics $G_p(s)$ between pulling speed and crystal radius $R_c(s)$ and the measurement dynamics $G_m(s)$ between the
crystal radius and the bright ring radius,
cf.\,Fig.\,\ref{fig:series-connection} also:
\begin{equation*}
	G(s) = \frac{R_{br}(s)}{V_p(s)} = \frac{R_{br}(s)}{R_c(s)} \cdot \frac{R_c(s)}{V_p(s)} = G_m(s) \cdot G_p(s).
\end{equation*}

For model I, one has
\begin{subequations}
\begin{align}
	G_{p,I}(s)&=\frac{-0.0047626}{(s-\SI{8.78e-6}{})\,(s+\SI{1.684e-3})} \label{Rc_by_Vp_I}\\
	G_{m,I}(s)&= -47.1864s+1.0041,\label{Rbr_by_Rc_I}
\end{align}
\end{subequations}
and for model II
\begin{subequations}
	\begin{align}
		G_{p,II}(s)&=\frac{-0.0048849}{(s-\SI{4.63e-7}{})\,(s-\SI{7.602e-6}{})} \label{Rc_by_Vp_II}\\
		G_{m,II}(s)&=\frac{R_{br}(s)}{R_c(s)} = -46.0050s+1.0043\label{Rbr_by_Rc_II}.
	\end{align}
\end{subequations}
The units are omitted here for the sake of clarity. The static gain of
\eqref{Rc_by_Vp_I} and \eqref{Rc_by_Vp_II} is in $\SI{}{\minute}$, while that of
\eqref{Rbr_by_Rc_I} and \eqref{Rbr_by_Rc_II} is dimensionless. The roots of the
numerator polynomial of a transfer function are called \emph{zeros}, the roots
of the denominator polynomial \emph{poles}. These roots are real-valued or occur
in complex conjugate pairs, since the coefficients of the polynomials are real.
Their unit is in $\SI{}{\radian\per\second}$. A system is stable if a finite
perturbation in any input signal results in a finite response in all output
signals. For a pole at $s=p_i$, the corresponding dynamics are stable if
$\Real(p_i) < 0$, i.e., if the pole $p_i$ is in the left half plane of the
complex plane. If $\Real(p_i) > 0$, the corresponding dynamics is unstable.
Unstable poles are often called \emph{Right Half Plane} (RHP) poles.  The system
is stable if all poles are in the left half plane. Clearly, stability (possibly
through control) is a basic requirement for operation of any system. The two models
differ in their pole configuration: Model I has one RHP pole at
$\SI{8.78e-6}{\radian\per\second}$, while in Model II both poles are unstable
with $p_1 = \SI{7.602e-6}{\radian\per\second}$ and $p_2 =
\SI{4.63e-7}{\radian\per\second}$.

A real-valued zero $s=z_i$ for which $\Real(z_i)>0$ is called a RHP zero, and
will cause an inverse response from the control input to the measurement. All
RHP zeros cause fundamental limitations in achievable control
performance\footnote{Also complex conjugate RHP zeros, which do not necessarily
cause an inverse response, see \citep{skogestad2007multivariable}.}. It is
apparent from \eqref{Rbr_by_Rc_I} and \eqref{Rbr_by_Rc_II} that both models have
a RHP zero at $\approx \SI{0.021}{\radian\per\second}$, as expected based on the
nonlinear simulations. The location of the right half-plane zero is nearly
independent of the heat transfer mechanism characterizing the heat flux into the
interface.

Finally, note that the transfer functions in \eqref{Rbr_by_Rc_I} and \eqref{Rbr_by_Rc_II}
are not physically realizable because they contain a direct differentiation of
the input signal. They only make sense when multiplied with the corresponding
terms in \eqref{Rc_by_Vp_I} and \eqref{Rc_by_Vp_II}, respectively.

\section{Design of a parallel compensator and feedback controller}\label{compensator_controller}
Using the linear models derived in Section\,\ref{sec:Linearization} a parallel compensator and stabilizing controller can be designed.

\begin{figure}[ht!]
	\centering
	\includegraphics[width=1.0\linewidth]{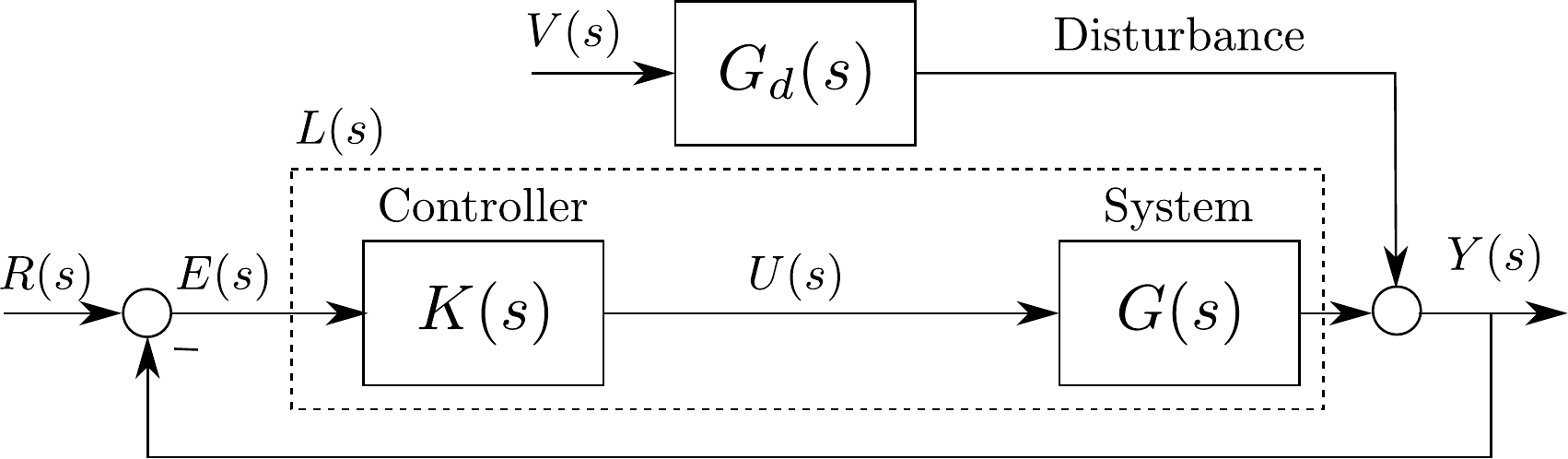}
	\caption{Illustration of a simple feedback control system with the transfer function $G(s)$ of the system to be controlled, the transfer function $K(s)$ of the controller and the transfer function of the disturbance model $G_d(s)$.}
	\label{fig:Feedback}
\end{figure}

\subsection{Basic feedback controller design} \label{sec:constraints:controller}
Feedback control is the most common type of control, illustrated in
Fig.\,\ref{fig:Feedback}.  This basic feedback schematic diagram shows a measurement $Y(s)$ being affected
by a disturbance $V(s)$ through the dynamics $G_d(s)$ and by the control input
$U(s)$ through the dynamics $G(s)$. The measurement signal $Y(s)$ is \emph{fed
back} and compared with its reference/ desired value $R(s)$ such that the
difference, alternatively called error $E(s)$, is used as an input to the
controller $K(s)$, which in turn calculates the control signal $U(s)$.

Feedback control is a remarkably powerful concept. It can stabilize unstable
systems, i.e. moving the unstable system poles to the left half of the complex
plane, and provide good performance, for example a quick response of the controlled quantity $y(t)$ to
changes in the reference $r(t)$, no or only limited overshooting of $y(t)$ or proper disturbance
rejection. However, it is not without caveats. In particular, poorly
designed feedback control may also cause instability, even for systems $G(s)$
which are stable on their own. Furthermore, and most important for this paper,
system zeros are unaffected by feedback.

\begin{figure}[ht!]
	\centering
	\includegraphics[width=1.0\linewidth]{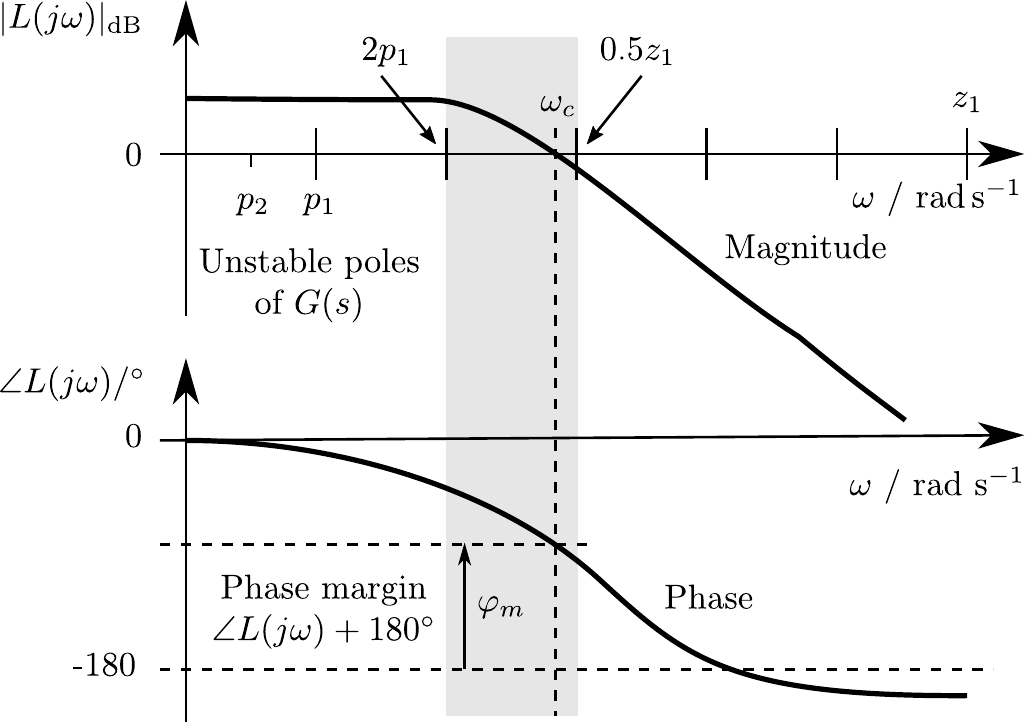}
	\caption{Illustrative and simplified sketch of a Bode diagram showing the frequency response of the open-loop system with $L(s) = K(s) \cdot G(s)$ and $s = j\omega$. The frequency axis is drawn in linear scale for sake of simplicity. The transfer function $G(s)$ of the system to be controlled is assumed to have two RHP poles denoted by $p_1, p_2$ (among other stable poles) and one RHP zero denoted by $z_1$. For a stable closed-loop system showing acceptable performance the crossover frequency $\omega_c$ should be roughly within the grey region and the phase margin $\varphi_m$ must be positive.}
	\label{fig:bode-general}
\end{figure}

A common way to determine suitable parameters of the controller is to utilize
the frequency response $L(j\omega) = K(j\omega) G(j\omega)$ of the so-called \emph{open-loop} system plotted in a Bode diagram. An illustrative example of a Bode diagram is shown
Fig.\,\ref{fig:bode-general}. The Bode diagram consists of two plots, the first
showing the magnitude $|L(j\omega)|$ in a logarithmic scale\footnote{dB or the base-10 logarithm are commonly used.} of the open-loop system depending on the
frequency $\omega$ and the second one showing the phase of $L(j\omega)$.

A well-designed feedback control system will have an open-loop frequency response with large magnitude at low frequencies, but small magnitude at high
frequencies.  Let $\omega_c$ denote the \emph{crossover frequency}, i.e., the
frequency where $|L(j\omega_c)|_\text{dB} = 0$. Assume that
$|L(j\omega)|_\text{dB} > 0 \; \forall \omega < \omega_c$, and
$|L(j\omega)|_\text{dB} < 0 \; \forall \omega > \omega_c$.  Then, the so-called
Bode stability criterion states that the \emph{phase margin} $\varphi_m =
180^{\circ} + \angle L(j\omega_c)$ \emph{must} be positive. Although some
sources present the Bode stability criterion only for open-loop stable systems,
it can also be applied to open-loop unstable systems -- as the Cz system under
discussion -- provided the number of open-loop unstable poles is known and the
steady-state phase is adjusted accordingly. Each RHP pole contributes a phase of $-180^{\circ}$ at $\omega = 0$. A small phase margin indicates that
the system may become unstable for a small error in the system model, and is
also an indication of poor performance, e.g.,\,large overshooting. Most control
engineers will insist on a phase margin of at least $45^{\circ}$.

Additional criterions can  be utilized to improve closed-loop performance: For
example, as a rule of thumb, $\omega_c$ should be about at least twice as large
as the \emph{fastest unstable pole} to ensure that any unstable dynamics is
properly `caught'. In case of non-minimum phase systems -- as discussed here --
another restriction comes into play: $\omega_c$ should be about less
than half the \emph{slowest zero} so the controller action is not dominated
by any inverse response. Clearly, the last two requirements show that the
presence of a RHP zero will introduce a typical conflict of objectives imposing
fundamental limitations on achievable performance for feedback control. An
extensive exposition of these issues can be found in
\citep{skogestad2007multivariable}, where further details, more precise
statements and theoretical justification can be found. Also refer to \ref{sec:control:frequencyanalysis} for an explanation of the term
\emph{non-minimum phase system}.

Coming back to the Cz system, for Model II, the required crossover frequency
will be determined mainly by the faster RHP pole (the one furthest from the
origin in the complex plane). The faster RHP pole
$7.602\mathrm{e}^{-6}\SI{}{\radian\per\second}$ in Model II is similar to the
RHP pole $8.78\mathrm{e}^{-6}\SI{}{\radian\per\second}$ of Model I. Therefore,
the control limitations of the two models are similar, even though the number of
unstable poles is different. Further details can be found in
\citep{BUKHARI2019129}.

\subsection{Basics of parallel compensator design} \label{sec:parallelcomp:basic}
The parallel compensator accounts for the anomalous behaviour of the measurement
signal, i.e., it moves RHP system zeros into the left plane only without altering the location of system
poles \citep{skogestad2007multivariable}.

\begin{figure}[ht!]
	\centering
	\includegraphics[width=\linewidth]{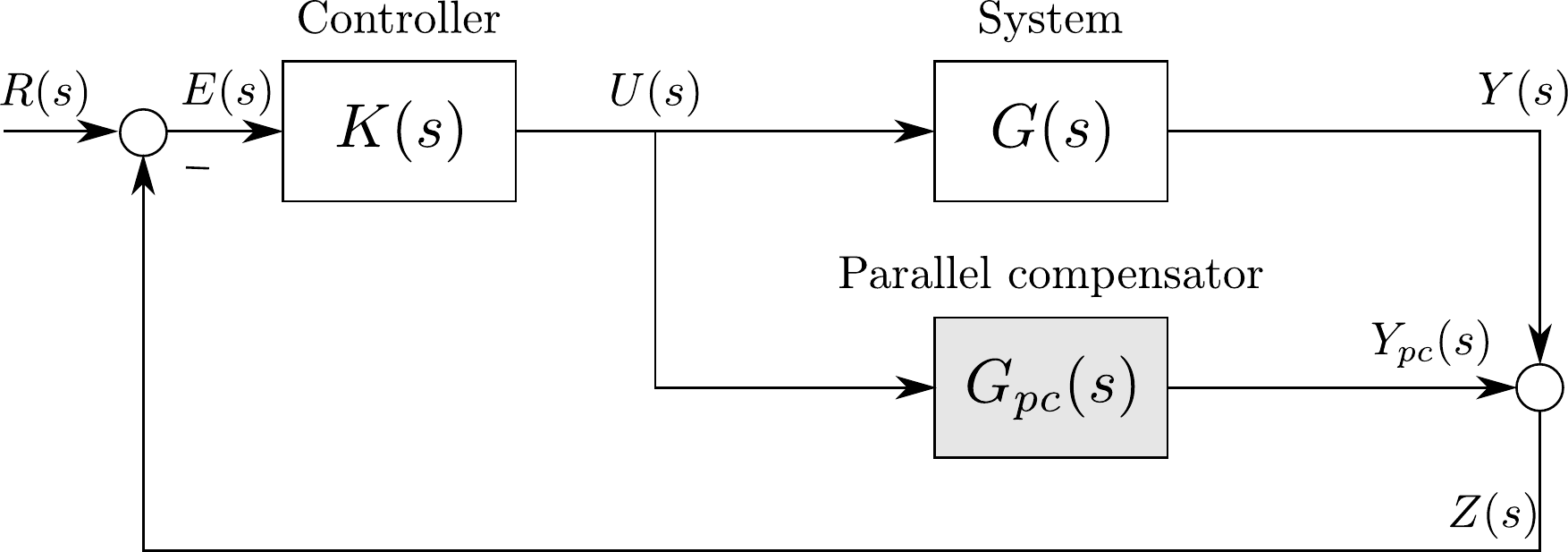}
	\caption{Feedback control combined with parallel compensation to remove performance limitations for feedback control caused by RHP zeros.}
	\label{fig:parallel}
\end{figure}

Several authors have proposed combining feedback control with
parallel compensation, thereby enabling the latter to remove the limitation in performance for feedback control. Such a schematic, with combined feedback and parallel compensator, is illustrated in Fig.\,\ref{fig:parallel}. Here, a parallel compensator $G_{pc}(s)$ is designed to ensure that the transfer function
from the input $U(s)$ to the compensated signal $Z(s) = Y(s) + Y_{pc}(s)$ does
not possesses any RHP zero to limit the performance of feedback control.

Unfortunately, in most physical systems with a control configuration such as
that depicted in Fig.\,\ref{fig:parallel}, the use of parallel compensation is not of much value.  This is because despite achieving good control of $Z(s)$,
the physical variable $Y(s)$, of interest, cannot be alleviated of the undesired effects of
the RHP zero(s).

However, as observed in Figs.\,\ref{fig:non_lin_conduction} and
\ref{fig:non_lin_convection}, the inverse response in the Cz process is merely
associated with the camera-based \emph{measurement} of the ingot radius, whereas the
\emph{actual crystal radius} is independent of the inverse behaviour. This opens an
opportunity for using parallel compensation to enable improved control of the
crystal radius $r_c$, even though the control of the measurement $r_{br}$ is not
improved.

\subsection{Compensator design} \label{sec:compensator:design}
Instead of using generic time-/ Laplace-domain symbols $r(t)/R(s)$,
$u(t)/U(s)$ and $y(t)/Y(s)$ as in Fig.\,\ref{fig:parallel} the symbols specific to the Cz
process will be used henceforth.

\begin{figure}[ht!]
	\centering
	\includegraphics[width=1.0\linewidth]{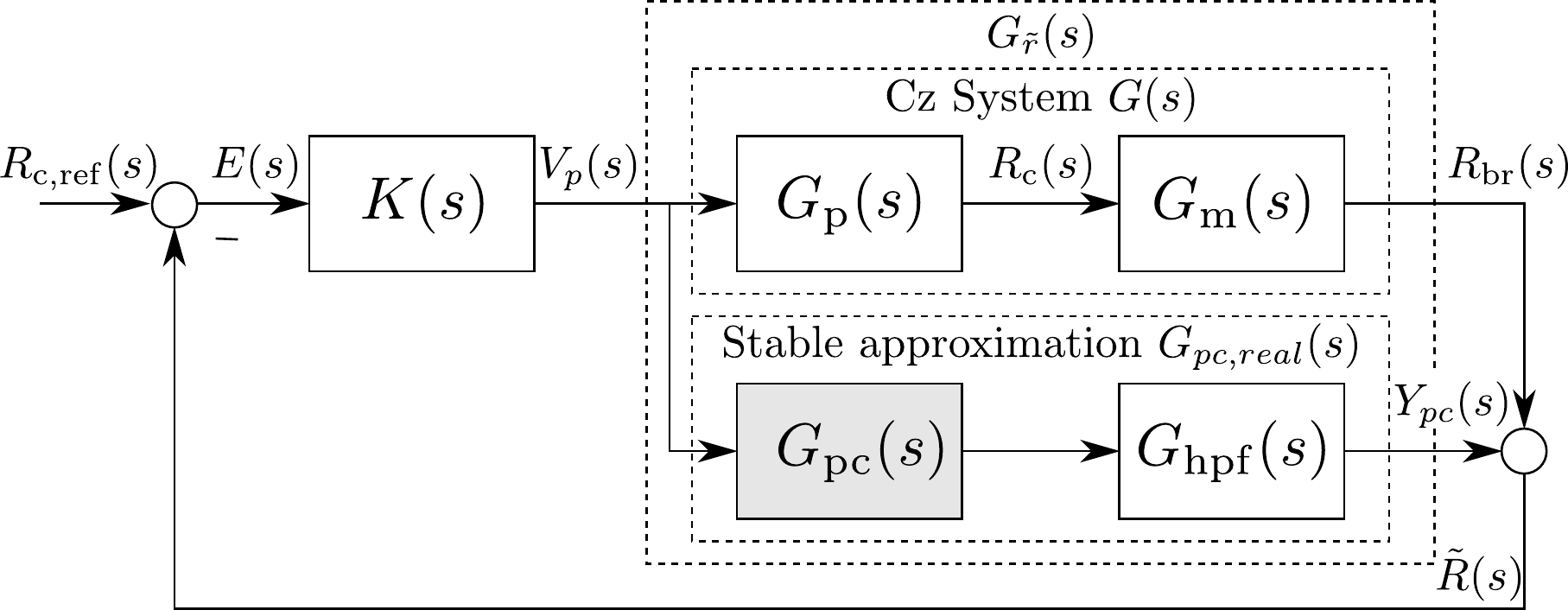}
	\caption{Basic block diagram of the feedback controlled Cz system split into the pure plant model $G_p(s)$ and the bright ring measurement model $G_m(s)$ together with a parallel compensator. The variable $R_{c,ref}$ defines the
	ingot radius reference and $\tilde{R}$ is the compensated radius
	measurement. The ideal parallel compensator $G_{pc}(s)$ is augmented by a high pass filter $G_{hpf}$ according to the procedure described in Section\,\ref{sec:compensator:design}. What is implemented in the real system is the stable approximation of $G_{pc}(s)G_{hpf}(s)$.}
	\label{fig:block_diagram}
\end{figure}

Fig.\,\ref{fig:block_diagram} shows a complete system block diagram that
includes both the compensator and the controller connected respectively in
parallel and cascade with the plant. The controller block marked as $K(s)$ is
the Automatic Diameter Control (ADC) as discussed in part I of this article
series. The parallel compensator used is a stable approximation of the ideal
compensator with transfer function $G_{pc}(s)$ augmented by a high pass filter
with transfer function $G_{hpf}(s)$. The need for both, the stable approximation
and the high pass filter, will be explained later. The following two design requirements need to be met:
\begin{enumerate}[labelindent=0ex,leftmargin=*]
	\item It is desired to keep the dynamics of the compensated measurement as
	close as possible to the one of the actual crystal radius, i.e.,
	$\tilde{R}(s) \approx R_c(s)$.
	\item The practices	in the Cz industry rely on using the camera-based
	measurement, followed by simply applying a bias to the measurement signal.
	By this approach they obtain the true \emph{steady-state} crystal radius
	from the camera measurement. The compensator design should allow the
	industry to continue applying the same bias to the compensated measurement
	$\tilde{R}(s)$. Hence, the compensator obtained in step 1 needs to be
	modified appropriately.
\end{enumerate}


Under these conditions the compensator -- based on model I -- can be designed as follows: Ignoring (for now) the high pass filter $G_{hpf}(s)$ in
Fig.\,\ref{fig:block_diagram}, the dependency of $\tilde{R}(s)$ on $V_p(s)$ gives
\begin{equation}
\tilde{R}(s) = \bigl(G_m(s)G_p(s)+G_{pc}(s) \bigr)V_p(s). \label{Eq:tf_comp}
\end{equation}

Hence, the ideal
parallel compensator can be derived as follows:
\begin{align}
	G_{\tilde r}(s) &\stackrel{!}{=} G_p(s) \nonumber\\
	G_p(s) G_m(s) + G_{pc}(s) &= G_p(s) \nonumber \\
	\Leftrightarrow \qquad G_{pc}(s) &= G_p(s)\left(1-G_m(s)\right) \label{Eq:comp_ideal}
\end{align}

It is apparent that the ideal parallel compensator as given in
\eqref{Eq:comp_ideal} would contain the same poles as the plant model $G_p(s)$,
including the unstable pole(s). Any system that consists of two parallel branches,
having identical unstable dynamics with common input and output, will not be
stabilizable by feedback as it will necessarily possess hidden unstable mode(s) \cite{skogestad2007multivariable}. It is, therefore, necessary to find a
stable approximation to the ideal parallel compensator $G_{pc}(s)$, i.e., an
approximation that removes the RHP zero, but does not destabilize the control loop.

In this case, the frequency of the RHP zero is higher than the frequencies
of the RHP pole(s) (cf.\,Section \ref{sec:Linearization}). Hence, the main interest is to have a good approximation of the unstable system at high frequencies to remove the effects of the RHP zero. The unstable dynamics are therefore suppressed by augmenting $G_{pc}(s)$ with a high pass filter $G_{hpf}(s)$, designed to cut off frequencies significantly below the frequency corresponding to the RHP zero. With the RHP zero at $\approx
\SI{0.021}{\radian\per\second}$ the transfer function $G_{hpf}(s)$ of the high
pass filter is chosen as
\begin{equation} \label{eq:hpf}
	G_{hpf}(s) = \frac{5000\,s}{5000\,s+1}
\end{equation}
with the cutoff frequency at $\SI{2e-4}{\radian\per\second}$.

Then, a stable/ unstable decomposition is performed on the augmented parallel
compensator. This means that the transfer function $G_{pc}(s) G_{hpf}(s)$ is
split into two transfer functions connected in parallel, one containing the stable dynamics, while the other, the unstable dynamics. The procedure is sketched in
\ref{sec:StableUnstable}. Since the unstable dynamics are slow, the
augmentation of the high pass filter makes the unstable part smaller compared to
the stable part. Hence, it is reasonable to use only the stable part of this
decomposition in the final implementation.

Finally, design requirement 2 has to be met. The stable part of the decomposition is therefore
adjusted to have zero steady-state gain such that $\tilde R = R_{br}$ in steady state. One ends up with the following  parallel compensator transfer function for model I
\begin{equation} \label{eq:pc:real}
	G_{pc,real}(s) = \frac{Y_{pc}(s)}{V_p(s)} = \frac{-0.22517s}{(s+\SI{1.684e-3}{})(s+\SI{2e-4}{})}.
\end{equation}
In the time domain, \eqref{eq:pc:real} is written as
\begin{align*}
	 \ddot{y}_{pc}(t)+\SI{1.884e-3}{}\,&\dot{y}_{pc}(t)\\
	 +\SI{3.368e-7}{}\,&y_{pc}(t)  =-0.22517\,\dot{v}_p(t)
\end{align*}
and this has to be implemented in the control system computer.


\subsection{Controller design}
The design of a stabilizing feedback controller follows the compensator design.
A PID controller is used  in this work because it is easy to implement in the
existing industrial control setup. The PID controller, represented by $K(s)$ in
a series/interacting form is given by
\begin{equation} \label{eq:PID}
K(s)=\frac{K_p\,(T_is+1)\,(T_ds+1)}{T_i s \left(1+\frac{T_ds}{N}\right)}.
\end{equation}
The tuned parameters for the PID controller and the resulting cross over
frequency are given in Table~\ref{tab:PID_params}.

\begin{table}[ht!]
	\centering
	\begin{tabular}{|c|c|}
		\hline
		Proportional gain ($K_p$) & \SI{0.01}{\per\second}\\
		\hline
		Integral Time const. ($T_i$)& \SI{5000}{\second}\\
		\hline
		Derivative Time const. ($T_d$)&\SI{650}{\second}\\
		\hline
		Filter coefficient ($N$)&100\\
		\hline
		Crossover frequency ($\omega_{c}$)& \SI{0.03275}{\radian\per\second}\\
		\hline		
	\end{tabular}
	\caption{Parameters of the PID controller $K(s)$.}
	\label{tab:PID_params}
\end{table}

\begin{figure}[ht!]
	\centering
	\includegraphics[width=1.0\linewidth]{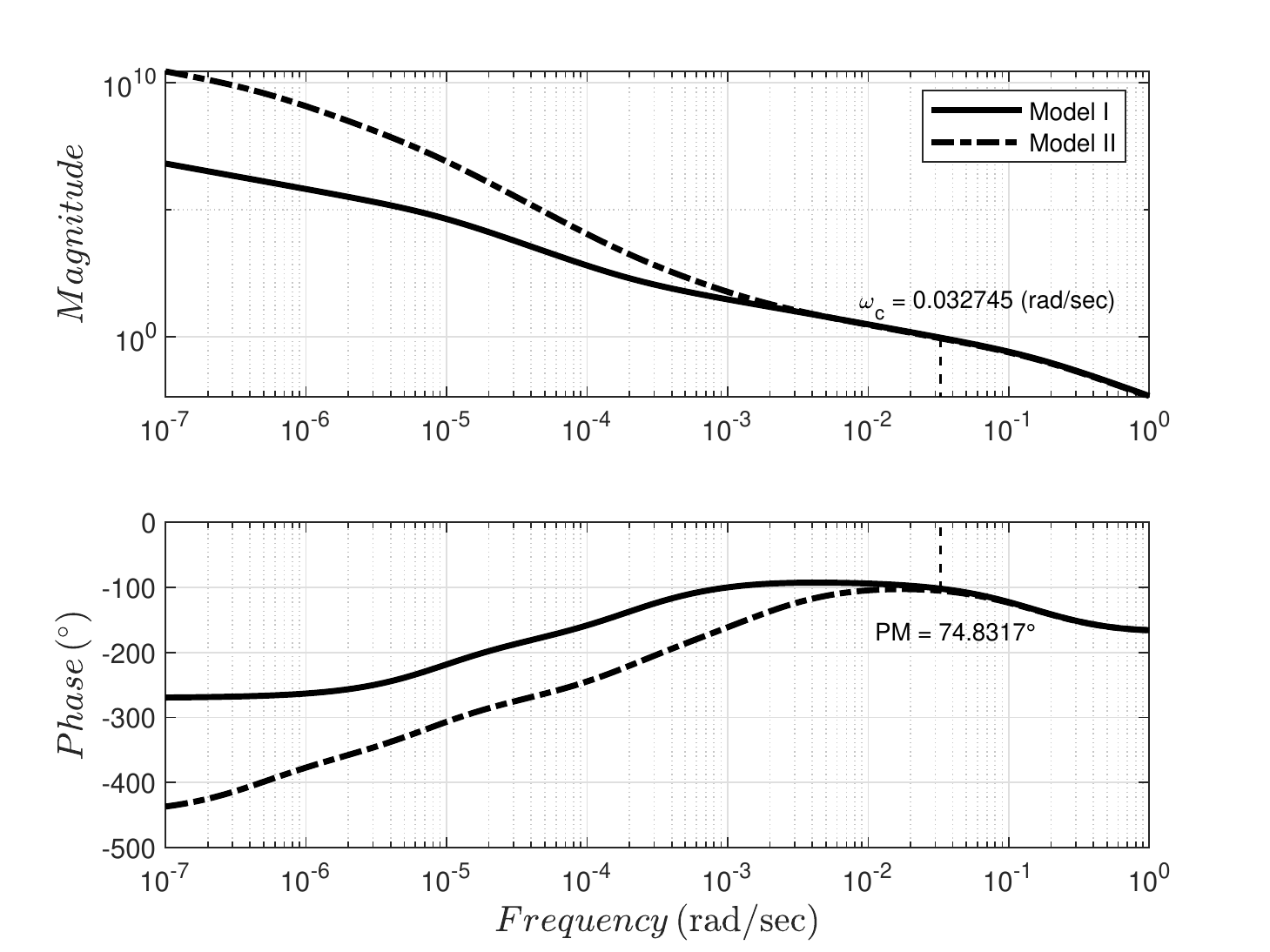}
	\caption{Frequency response of the compensated plants (models I and II) with a feedback controller.}
	\label{fig:frequency_responses}
\end{figure}

The achieved crossover frequency $\omega_{c}\approx$
\SI{0.03275}{\radian\per\second} is much higher than half of the zero at
$\approx$ \SI{0.02}{\radian\per\second} which would have been the limiting
factor in control design without parallel compensation. So the limitations
imposed by the RHP zero is quite clearly mitigated and a higher system bandwidth is
achieved.  The frequency response in Fig.\,\ref{fig:frequency_responses} shows
that the proposed PID controller with the compensator stabilizes both models, as the phase margin of the compensated plant is around $74^{\circ}$ for both models.

\subsection{Closed-loop performance}

\begin{figure} [ht!]
	\centering
	\includegraphics[width=1.0\linewidth]{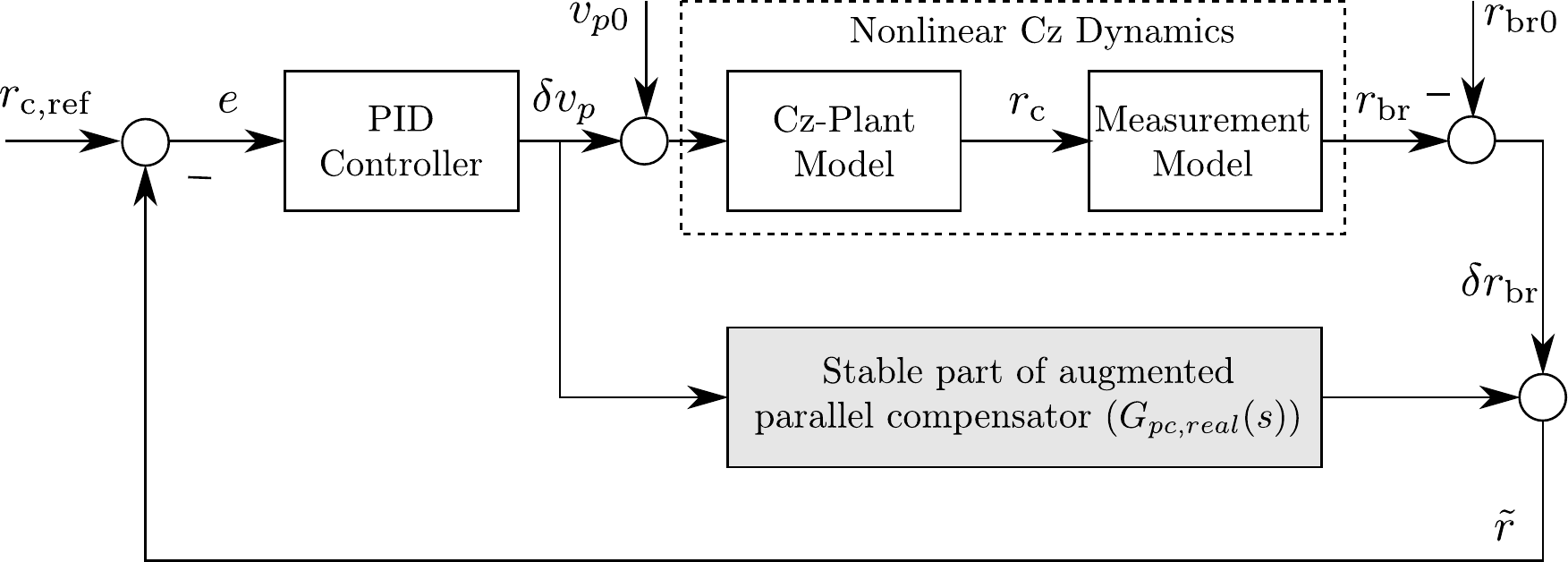}
	\caption{Block diagram for the closed-loop testing of nonlinear Cz growth dynamics.}
	\label{fig:compensator_block_diagram_v4} 	
\end{figure}

The closed-loop testing of nonlinear Cz growth dynamics, in the presence of both
parallel compensator as well as the feedback controller ($K$), is schematically
illustrated in Fig.\,\ref{fig:compensator_block_diagram_v4}.

In Fig.\,\ref{fig:compensator_block_diagram_v4}, $\delta\,v_p$ and
$\delta\,r_{br}$ are the deviation variables, whereas $v_{po}$ and $r_{br0}$ are
the steady-state values (points at which linearization was performed) such that
the input to and the output from the nonlinear Cz dynamics are $v_p\, =\,
v_{po}\, +\, \delta\,v_p$ and $R_{br}\, = \,r_{br0}\,+\,\delta\,r_{br}$,
respectively.

In order to determine the extent to which the nonlinearities in the Cz system may be excited, the responses to two different crystal radius reference
trajectories are simulated.

\begin{figure} [ht!]
	\centering
	\includegraphics[width=1.0\linewidth]{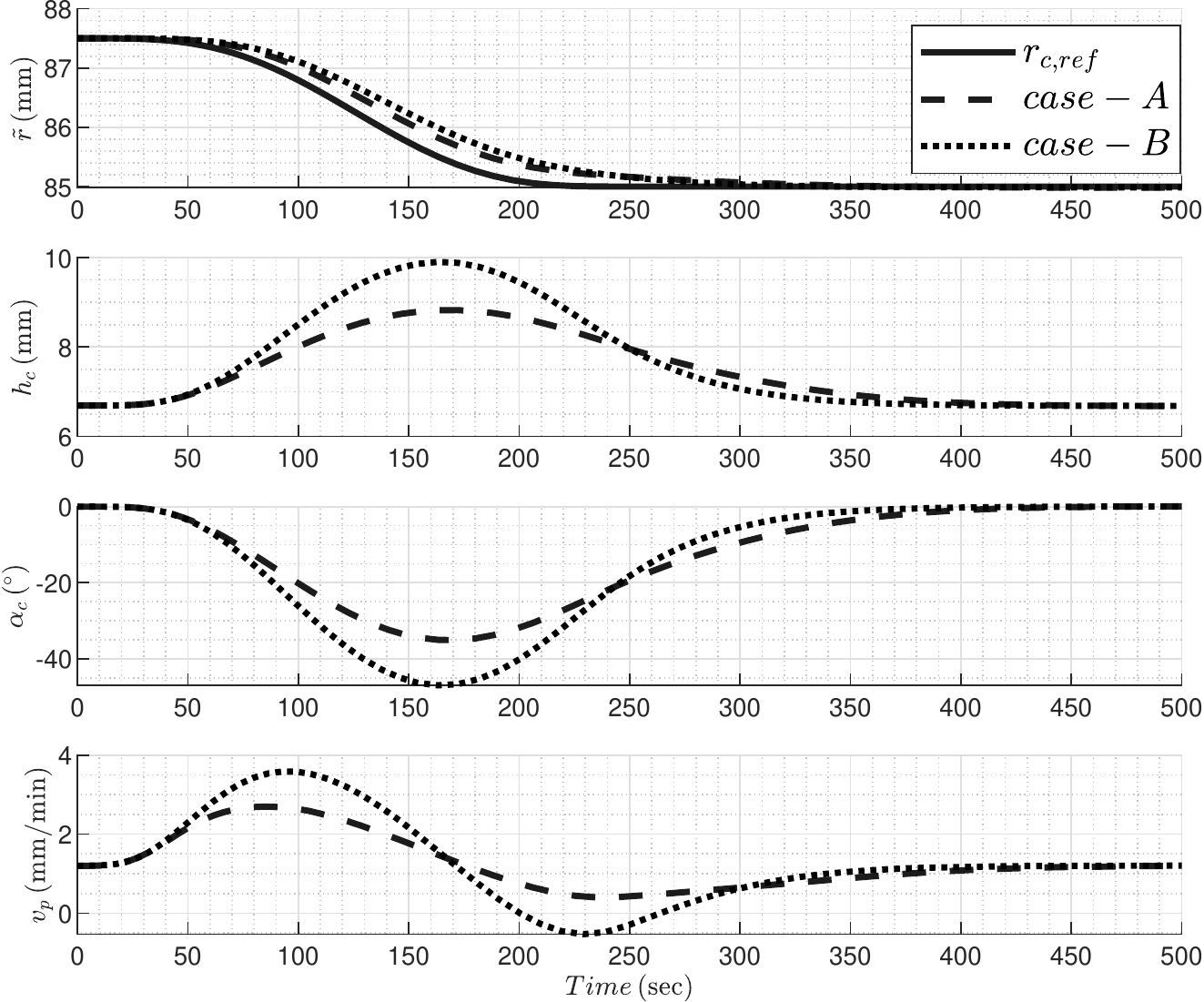}
	\caption{Comparison of system responses to $r_{c,ref}$ with large amplitude change (case-A) and small amplitude change (case-B: \emph{responses are scaled up by a factor of 5}). Top row: compensated measurement $\tilde{r}$, second row: meniscus height $h_c$, third row: cone angle $\alpha_c$, last row: control input $v_p$.}
	\label{fig:nlanalysis} 	
\end{figure}

In the first case, referred to as (\textbf{case-A}) in
Fig.\,\ref{fig:nlanalysis}, the response to a smooth reduction in the crystal
radius reference of \SI{2.5}{\milli\meter} is simulated. In the second case
(\textbf{case-B}), a scaled version of the same smooth crystal radius reference
trajectory is applied, changing the reference by \SI{0.5}{\milli\meter}. In
Fig.\,\ref{fig:nlanalysis}, the responses of \textbf{case-B} are scaled by a
factor of 5 to make them easily comparable to the responses of \textbf{case-A}.
Fig.\,\ref{fig:nlanalysis} shows that the system with the proposed control is
relatively insensitive to nonlinearities for smooth reference changes of
reasonable magnitude.

\section{Responses to temperature disturbances }\label{sec:temp_model}
Hitherto, the simulated growth rate variations did not take temperature dynamics
into account. However, temperature variations are a major source of disturbances
to the crystal growth rate and thereby also to the crystal radius control.  To
assess control performance in the presence of growth rate variations caused by
temperature disturbances, the overall Cz model (with both growth and temperature
dynamics) needs to be used. A qualitative heater model augmented with the
crystal growth dynamics has been presented in Section\,3 of the preceding
article \citep{BUKHARI2020A}.  Thus, in the following, the overall Cz dynamics
(both the growth model as well as heater model) are under the combined influence
of the two system inputs, i.e., pulling speed $v_p$ and heater input $Q_H$.

\subsection{Temperature effects on the overall system performance}
 During a typical growth cycle in the body stage, a feedforward temperature
 (target temperature) trajectory is applied to the temperature controller to
 compensate for the slow temperature dynamics. In an actual process, the
 temperature feedforward trajectory has an increasing trend (typically in the
 range of $\approx$ \SI{0.1}{\kelvin\per\minute}) to compensate for the
 following phenomena occurring throughout the growth cycle within the Cz growth
 chamber:
  \begin{itemize}[labelindent=1ex,leftmargin=*]
 	\item A gradual uplift of the crucible, therefore progressively reducing the crucible exposure to the heaters.
 	\item With the ongoing crystallization, the crystal continues to protrude into the colder areas above the heat shield, thereby increasing the heat transfer away from the interface.
 \end{itemize}

The perfect temperature feedforward trajectory is hard to establish, due to effects such as aging and continual replacement of components in the hot zone, variations between pullers, etc.  An imperfect temperature trajectory will act as a disturbance to the crystal growth rate and hence also affects the crystal radius 
control.  Note that the heater model in this work is qualitative, and therefore
does not include the afore-mentioned phenomena causing a need for an increasing
feedforward trajectory under actual growth conditions.  However, effects of an
imperfect temperature trajectory can be simulated\footnote{High accuracy not
claimed here though.}, since changes to the temperature controller reference
will affect the melt temperature in the model and hence also affect the crystal
growth rate. That is, in our simulation on the simplified model, the temperature
feedforward trajectory does not represent the actual feedforward trajectory, but
rather the error in the feedforward trajectory in an actual plant.

\begin{figure} [ht!]
	\centering
	\includegraphics[width=1.0\linewidth]{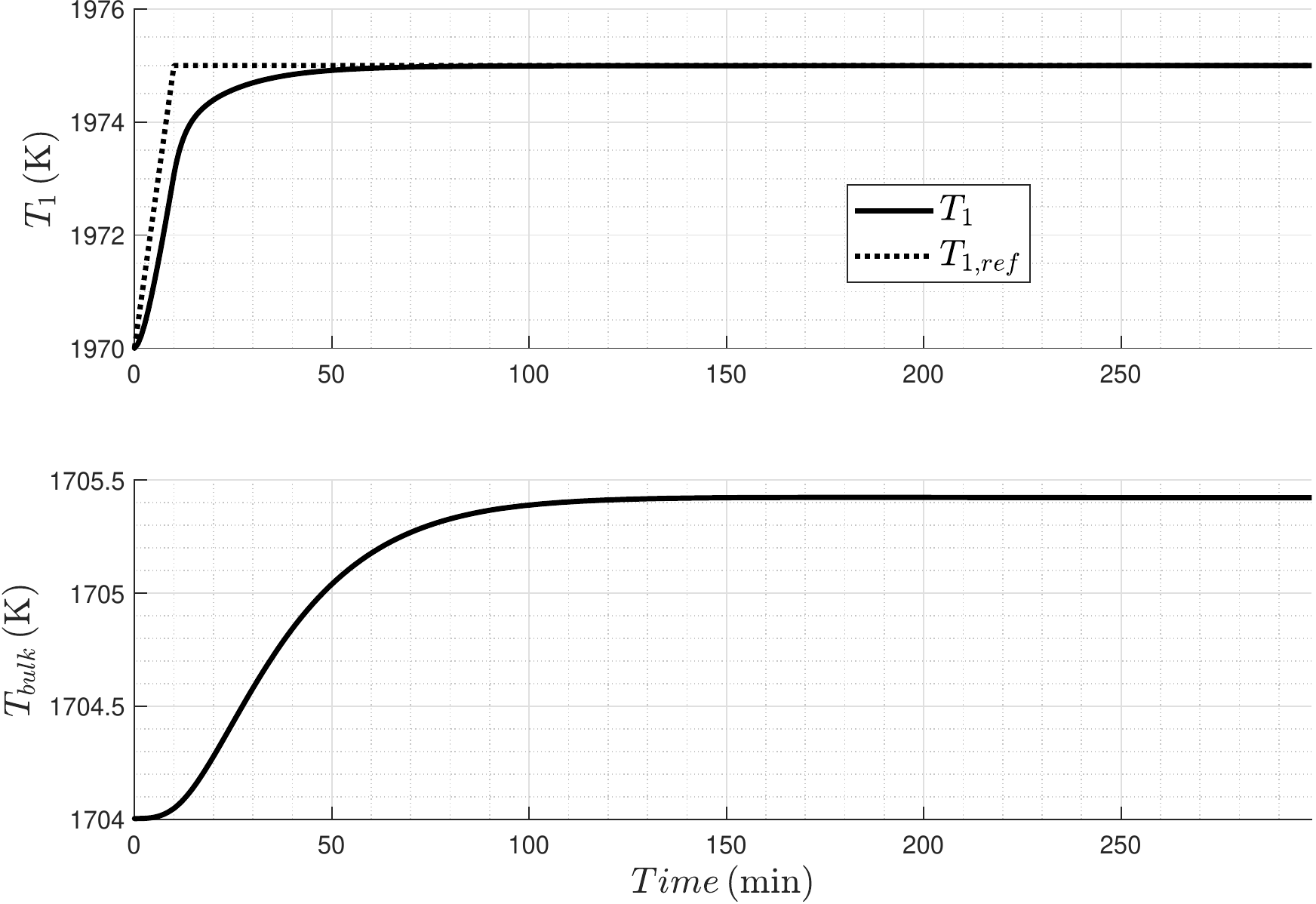}
	\caption{Temperature responses for model I. Top pane: temperature sensed by the pyrometer $T_1$ versus the target temperature trajectory $T_{1,ref}$. Bottom pane: temperature of the bulk $T_{bulk}$ of the melt contained within the crucible.}
	\label{fig:MH1_tempresponse_Td650Kx1_tempPID} 	
\end{figure}

To investigate the effects of temperature disturbances on crystal radius
control, the reference for temperature $T_1$ (sensed by pyrometer) is increased
linearly (at the rate of \SI{0.5}{\kelvin\per\minute}) to simulate the variation
in the target temperature trajectory.

Compared to the typical difference between the actual and the ideal feedforward trajectory, the simulated feedforward trajectory must be considered to represent
a rather strong disturbance. The resultant responses for temperatures in
different lumped volumes (cf. lumped heater model given in \cite{BUKHARI2020A})
is shown in Fig.\,\ref{fig:MH1_tempresponse_Td650Kx1_tempPID}. These temperature
changes, in turn, affect the radii responses ($r_c$, $r_{br}$, $\tilde{r}$) via
variation in growth rate $v_g$.

\begin{figure} [ht!]
	\centering
	\includegraphics[width=1.0\linewidth]{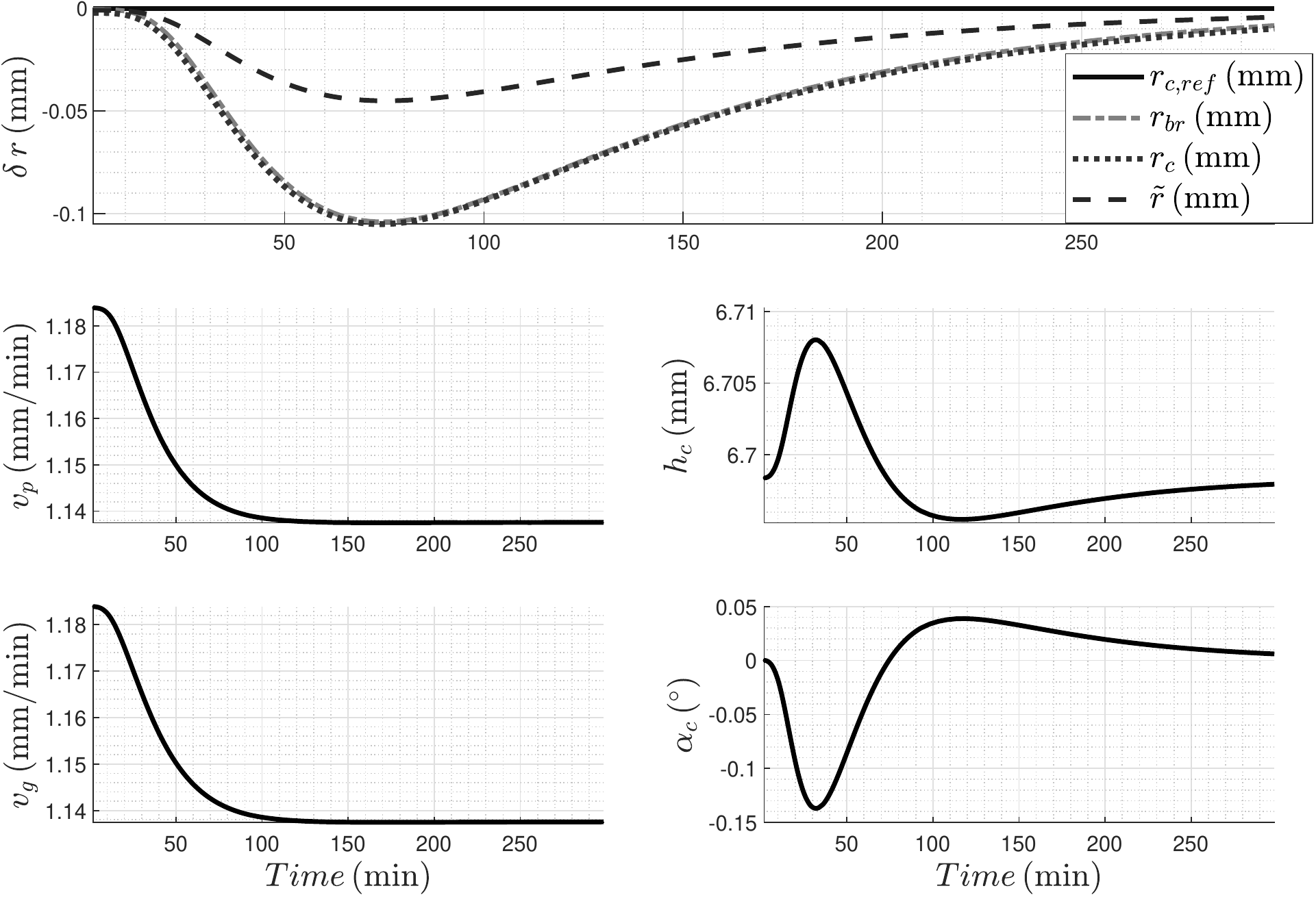}
	\caption{Closed-loop responses for model I with a change in temperature set-point trajectory shown in Fig.\,\ref{fig:MH1_tempresponse_Td650Kx1_tempPID}. Top row: crystal radius $r_c$, compensated measurement $\tilde{r}$ and bright ring radius measurement $r_{br}$ versus constant $r_{c,ref}$. Middle row (left $\rightarrow$ right): control input $v_p$ and meniscus height $h_c$. Bottom row (left $\rightarrow$ right): growth rate $v_g$ and growth angle $\alpha_c$.}
	\label{fig:MH1_response_Td650Kx1_tempPID} 	
\end{figure}

\begin{figure} [ht!]
	\centering
	\includegraphics[width=1.0\linewidth]{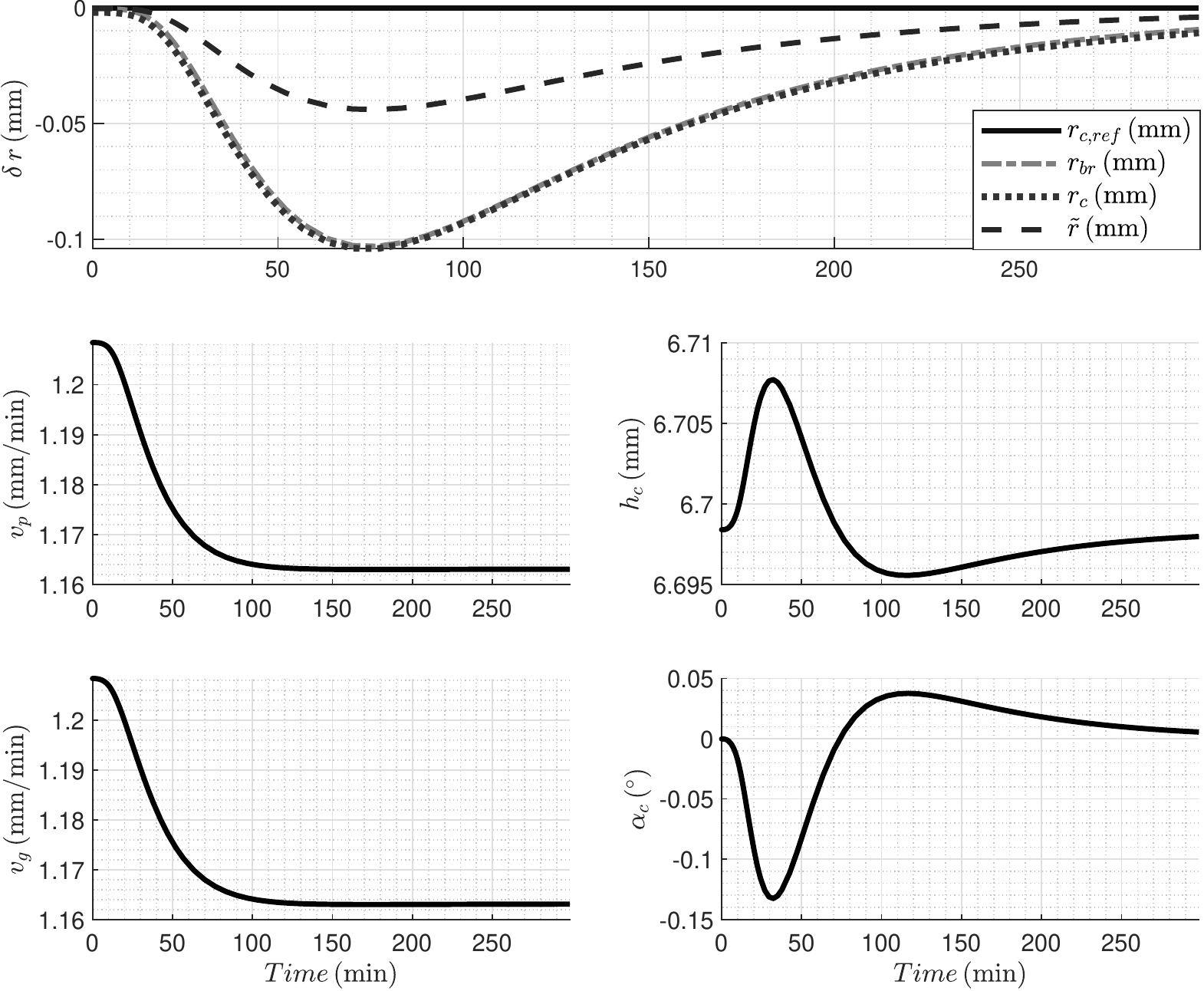}
		\caption{Closed-loop responses for model II with a change in temperature set-point trajectory shown in Fig.\,\ref{fig:MH1_tempresponse_Td650Kx1_tempPID}. Top row: crystal radius $r_c$, compensated measurement $\tilde{r}$ and bright ring radius measurement $r_{br}$ versus constant $r_{c,ref}$. Middle row (left $\rightarrow$ right): control input $v_p$ and meniscus height $h_c$. Bottom row (left $\rightarrow$ right): growth rate $v_g$ and growth angle $\alpha_c$.}
	\label{fig:MH2_response_Td650Kx1_tempPID} 	
\end{figure}

For the given change in heater set-point trajectory as depicted in top pane of
Fig.\,\ref{fig:MH1_tempresponse_Td650Kx1_tempPID}, the resulting system responses
at the crystallization growth interface are shown in
Figs.\,\ref{fig:MH1_response_Td650Kx1_tempPID} and
\ref{fig:MH2_response_Td650Kx1_tempPID} for models I and II, respectively.

It is apparent from the radii responses (cf.
Figs.\,\ref{fig:MH1_response_Td650Kx1_tempPID} and
\ref{fig:MH2_response_Td650Kx1_tempPID}) that the designed controller and
parallel compensator have successfully overcome the temperature variations
leading to a change in the growth rate, and that the resulting variation in
crystal radius is small.

\section{Conclusions}
In continuation of the investigation of the inverse response in the preceding
article of this two-article series, this work focuses on controller design for
crystal radius control. The proposed control involves compensation for the
inverse response behaviour with the use of a parallel compensator, and a
conventional PID controller for stabilization and disturbance rejection. The
performance of the resulting system is assessed using nonlinear simulations
including temperature disturbances that alter the growth rate. The results show
that the designed controller stabilizes the crystal radius and has satisfactory
disturbance rejection capabilities for the disturbances that typically occur in
the Cz process.

\appendix

\section{Some control engineering background} \label{sec:control}
This appendix will introduce some background to control engineering basics, aimed at making
the contribution of this paper accessible to people outside the control
community. The descriptions will necessarily be imprecise and incomplete, aimed
at conveying the main ideas. Readers desiring more information are referred to
standard introductory textbooks on control, and to
\citep{skogestad2007multivariable}.

\subsection{Time domain and transfer functions models} \label{sec:control:models}
Control design and analysis is based on a dynamical model of the system
considered.  The dynamical model may be in either of the following two types:

\paragraph{Time domain model.} In this model, the system behaviour is modelled
using differential equations, most often ordinary differential equations.
Modeling based on first principles (physical and chemical laws/principles)
results in this type of model. The resulting models are generally of the form:
\begin{equation}
    \dot{\mathbf{x}} = \mathbf{f}(\mathbf{x},\mathbf{u}) \qquad \qquad \mathbf{y} = \mathbf{g}(\mathbf{x},\mathbf{u})
\end{equation}
where the variables in the vector $\mathbf{x}$ are called \emph{state
variables} (typically related to some conserved entity for some control volume,
i.e., mass, energy, temperature, etc.), $\dot{\mathbf{x}}$ is the time
derivative of $\mathbf{x}$, $\mathbf{u}$ denotes the input(s) to the system, and
$\mathbf{y}$ is the measurement(s). To simplify analysis and design, the time domain model is often \emph{linearized} around some conditions of interest,
usually a desired steady-state operating point. Linearization involves deriving
a Taylor series expansion of the model with respect to $\mathbf{x}$ and
$\mathbf{u}$, and terminating the expansion after the first order terms.  If the
linearization is performed around a steady state, the constant (zero order) term
of the Taylor series will be zero, and one is left with the linearized model,
i.e., the first order term from the Taylor series.  With the linearized model,
presenting a linear approximation of the system dynamics, deviated by a small
neighborhood around the operating point, its mathematical representation in
terms of deviation variables is given as:
\begin{equation} \label{eq:LinSys:General}
\delta\dot{\mathbf{x}} = \mathbf{A} \cdot \delta \mathbf{x} + \mathbf{B}\cdot \delta \mathbf{u} \qquad
\delta \mathbf{y} = \mathbf{C}\cdot \delta \mathbf{x} + \mathbf{D}\cdot \delta \mathbf{u}
\end{equation}
with constant matrices $\mathbf{A},\mathbf{B},\mathbf{C},\mathbf{D}$ of
appropriate dimensions. System \eqref{eq:LinSys:General} is stable if all eigenvalues of $\mathbf{A}$ have a negative real part. It is noteworthy that the variables $\delta
\mathbf{x}$, $\delta \mathbf{y}$, $\delta \mathbf{u}$ in the linearized model are
\emph{deviation variables}, representing the difference from the operating
conditions.

\paragraph{Transfer function model.}  Using the so-called \emph{Laplace
transform}, the linearized model can be transformed to an equivalent model in
the Laplace domain, resulting in a transfer function model.  The transfer
function model describes the dynamic effects of the input on the output in terms
of the complex-valued Laplace variable $s$.  For a system with one input and one
output (measurement), the transfer function model can be expressed as
\begin{multline} \label{eq:TransferFunc:General}
    Y(s) = \frac{N(s)}{M(s)} U(s) = \\ \left( \mathbf{C}(sI-\mathbf{A})^{-1}\mathbf{B}+\mathbf{D} \right) U(s) = G(s)U(s),
\end{multline}
where $N(s)$ and $M(s)$ are polynomials in the Laplace variable $s$, with real
coefficients and $\mathbf{A}$, $\mathbf{B}$, $\mathbf{C}$, $\mathbf{D}$ from
\eqref{eq:LinSys:General}. Throughout this paper, it is assumed that any common
terms in $N(s)$ and $M(s)$ have been cancelled, as such terms represent dynamics that do not affect the input-output behaviour of the system.

\subsection{Frequency analysis and non-minimum phase systems} \label{sec:control:frequencyanalysis}
\paragraph{Frequency analysis} Frequency analysis involves evaluating the
transfer function $G(s)$ for $s = j\omega$, where $j = \sqrt{-1}$ and $\omega$
is real (and usually nonnegative). The results are often presented using the
transfer function \emph{magnitude} $|G(j\omega)|$ and \emph{phase} $\angle
G(j\omega)= \tan^{-1} \frac{Im(G(j\omega))}{Re(G(j\omega))}$. The frequency
analysis describes the stationary response of the system to a sinusoidal input
oscillating at frequency $\omega = \frac{2\pi}{t_p}$, where $t_p$ is the period
for one complete oscillation. Due to the linearity of the system, a sinusoidal input $U(j\omega)$ will cause a sinusoidal response in the output $Y(j\omega)$ with the same frequency $\omega$. The transfer function magnitude $|G(j\omega)|$ describes the amplification through $G(s)$, i.e., $|G(j\omega)|$ =$ \frac{|Y(j\omega)|}{|U(j\omega)|}$, whereas the phase $\angle
G(j\omega)$ describes the time shift between the input and output oscillation. Most often the oscillation in the
output will lag behind the oscillation in the input -- which corresponds to a
negative phase.  If the output lags the input by a full oscillation period, this
corresponds to a phase of $-2\pi$ radians (= -360$^\circ$).

\paragraph{Non-minimum phase system}  A major source of difficulty in control
design is that the magnitude and phase plots in the Bode diagram are not
independent. One goal of frequency-domain control design is to achieve a large
loop gain at low frequencies and a small one at high frequencies. The more
quickly the transfer function magnitude decreases with increasing frequency, the
more negative the transfer function phase will have to be. Considering the Bode
stability criterion, the magnitude plot cannot be very steep in the region
around the crossover frequency. For the magnitude plot of any physically realizable transfer function, there is a corresponding minimum phase curve. Non-minimum phase
systems are systems with more negative phase than what could be possible given
the magnitude plot, and considering the Bode stability criterion, it is clear
that such systems are therefore more difficult to control. Of particular concern
here is the non-minimum phase effect of RHP zeros, since they increase the
magnitude while making the phase more negative.


\subsection{Stable-unstable decomposition of a transfer function} \label{sec:StableUnstable}
The system \eqref{eq:LinSys:General} with a single input $u$ and a single output $y$ can be written as
\begin{equation} \label{eq:LinSys:BeforeDecomp}
	\begin{pmatrix}
		\delta\dot{\mathbf{x}}\\
		\delta y	
	\end{pmatrix}
	=
	\begin{pmatrix}
		\mathbf{A} &\mathbf{B}\\
		\mathbf{C} &d
	\end{pmatrix}
	\cdot
	\begin{pmatrix}
		\delta \mathbf{x}\\
		\delta u
	\end{pmatrix}
\end{equation}
with $\mathbf{A} \in \mathbb{R}^{n \times n}$, $\mathbf{B} \in \mathbb{R}^{n \times 1}$, $\mathbf{C} \in \mathbb{R}^{1 \times n}$ and $d \in \mathbb{R}$. By applying the \emph{Schur decomposition}, it is always possible to decompose \eqref{eq:LinSys:BeforeDecomp} into the following form
\begin{equation} \label{eq:LinSys:SchurDecomp}
	\begin{pmatrix}
		\delta\dot{\mathbf{x}}_1\\
		\delta\dot{\mathbf{x}}_2\\
		\delta y	
	\end{pmatrix}
	=
	\underbrace{
		\begin{pmatrix}
			\mathbf{A}_{11} &\mathbf{A}_{12} &\mathbf{B}_1\\
			0               &\mathbf{A}_{22} &\mathbf{B}_2\\
			\mathbf{C}_1    &\mathbf{C}_2    &d
		\end{pmatrix}
	}_{= A^\ast}
	\cdot
	\begin{pmatrix}
		\delta \mathbf{x}_1\\
		\delta \mathbf{x}_2\\
		\delta u
	\end{pmatrix}
\end{equation}
with $\delta x_1 \in \mathbb{R}^{n_1}$, $\delta x_2 \in \mathbb{R}^{n_2}$,  $n_1 + n_2 = n$ and the corresponding dimensions of $\mathbf{A}_{11}$, $\mathbf{A}_{12}$, $\mathbf{A}_{22}$, $\mathbf{B}_1$, $\mathbf{B}_2$, $\mathbf{C}_1$ and $\mathbf{C}_2$. System \eqref{eq:LinSys:SchurDecomp} has the special property that all eigenvalues of $\mathbf{A}_{11}$ have a negative real part while all eigenvalues of $\mathbf{A}_{22}$ have a nonnegative real part. This means that the subsystem with states $\delta \mathbf{x}_1$ is stable while the subsystem with states $\delta \mathbf{x}_2$ is unstable. To get the transfer functions $G_\mathrm{stable}(s)$, $G_\mathrm{unstable}(s)$ representing the stable and the unstable part of \eqref{eq:LinSys:BeforeDecomp}, respectively, it is necessary to transform \eqref{eq:LinSys:SchurDecomp} into the following form:
\begin{equation} \label{eq:LinSys:Decoup}
	\begin{pmatrix}
		\delta\dot{\bar{\mathbf{x}}}_1\\
		\delta\dot{\bar{\mathbf{x}}}_2\\
		\delta y	
	\end{pmatrix}
	=
	\underbrace{
		\begin{pmatrix}
			\mathbf{A}_{11} &0              &\mathbf{B}_1 - \mathbf{Q} \mathbf{B}_2\\
			0               &\mathbf{A}_{22}                           &\mathbf{B}_2\\
			\mathbf{C}_1    &\mathbf{C}_1 \mathbf{Q} + \mathbf{C}_2    &d
		\end{pmatrix}
	}_{\bar A^\ast}
	\cdot
	\begin{pmatrix}
		\delta \bar{\mathbf{x}}_1\\
		\delta \bar{\mathbf{x}}_2\\
		\delta u
	\end{pmatrix}
\end{equation}
with a transformation
\begin{equation} \label{eq:Transformation}
	\begin{pmatrix}
		\delta \mathbf{x}_1\\
		\delta \mathbf{x}_2
	\end{pmatrix}
	=
	\underbrace{
		\begin{pmatrix}
			\mathbf{I}_n &\mathbf{Q}\\
			\mathbf{0}   &\mathbf{I}_m
		\end{pmatrix}
	}_{=: \mathbf{T}}
	\cdot
	\begin{pmatrix}
		\delta \bar{\mathbf{x}}_1\\
		\delta \bar{\mathbf{x}}_2
	\end{pmatrix}.
\end{equation}
Here $\mathbf{I}_n$, $\mathbf{I}_m$ are $n \times n$ and $m \times m$ identity matrices, respectively, and $\mathbf{Q}$ is a $n \times m$ matrix which needs to be determined to get a system of structure \eqref{eq:LinSys:Decoup}. From \eqref{eq:LinSys:Decoup} the output $Y$ in the Laplace domain can be computed as (cf.\,\eqref{eq:TransferFunc:General})
\begin{multline}
	Y(s) = \left( \underbrace{\mathbf{c_1}(sI-\mathbf{A}_{11})^{-1}(\mathbf{B}_1 - \mathbf{Q} \mathbf{B}_2) + \frac{d}{2}}_{G_\mathrm{stable}(s)}\right) U(s) + \\ \left( \underbrace{(\mathbf{C}_1 \mathbf{Q} + \mathbf{C}_2)(sI-\mathbf{A}_{22})^{-1}\mathbf{B}_2+\frac{d}{2}}_{G_\mathrm{unstable}(s)} \right) U(s).
\end{multline}
The matrix $\mathbf{Q}$ required for the transformation from \eqref{eq:LinSys:SchurDecomp} to \eqref{eq:LinSys:Decoup} can be computed as follows: Applying \eqref{eq:Transformation} to \eqref{eq:LinSys:SchurDecomp} one obtains the matrix
\begin{multline}
	\mathbf{T}^{-1} \mathbf{A^\ast} \mathbf{T} =\\
	\begin{pmatrix}
		\mathbf{A}_{11} &\mathbf{A}_{11}\mathbf{Q} + \mathbf{A}_{12} - \mathbf{Q} \mathbf{A}_{22} &\mathbf{B}_1 - \mathbf{Q} \mathbf{B}_2\\
		0 &\mathbf{A}_{22} &\mathbf{B}_2\\
		\mathbf{C}_1    &\mathbf{C}_1 \mathbf{Q} + \mathbf{C}_2    &d
	\end{pmatrix}
\end{multline}
which must be equal to $\bar{\mathbf{A}}^\ast$. This results in the condition $\mathbf{A}_{11}\mathbf{Q} + \mathbf{A}_{12} - \mathbf{Q} \mathbf{A}_{22} = 0$ which is equal to solving the well-known \emph{Sylvester equation} $A X + X B = C$ with $A = \mathbf{A}_{11}$, $X = \mathbf{Q}$, $B = -\mathbf{A}_{22}$ and $C = -\mathbf{A}_{12}$. This equation will always have a unique solution $\mathbf{Q}$ since $\mathbf{A}_{11}$ and $\mathbf{A}_{22}$ do not have common eigenvalues due to the Schur decomposition mentioned above.

\section*{Acknowledgements}
This work has been funded by the Norwegian Research Council's ASICO project No. 256806/O20.

\bibliographystyle{plain}
\bibliography{czochralski_final_MH}	

\end{document}